  \documentclass[twocolumn,pra,showpacs,floatfix]{revtex4}
\usepackage{graphicx}
\usepackage{subfigure}
\usepackage[]{amsmath}
\usepackage{dcolumn}
\usepackage{bm}
\usepackage{tabularx}

\graphicspath{.}

\usepackage{color}

\begin{document}

\title{ Complete Active Space
 multiconfiguration Dirac-Hartree-Fock calculations
 of hyperfine structure constants of the gold atom}

\affiliation{Instytut Fizyki imienia~Mariana~Smoluchowskiego,
             Uniwersytet Jagiello{\'n}ski \\
             Reymonta~4, 30-059~Krak{\'o}w, Poland \\
             \email{Bieron@uj.edu.pl}}
\affiliation{National Institute of Standards and Technology \\
             Gaithersburg, MD 20899-8420, USA}
\affiliation{Laboratoire Kastler Brossel,
\'Ecole Normale Sup\' erieure \\
 CNRS; Universit\' e P.~et M.~Curie - Paris 6 \\
Case 74; 4, place Jussieu, 75252 Paris CEDEX 05, France}
\affiliation{Nature, Environment, Society \\
             Malm\"o University, S-205~06 Malm{\"o}, Sweden}
\affiliation{Department of Chemistry, University of Helsinki \\
             POB 55 (A.I.~Virtasen aukio 1), 00014 Helsinki, Finland}

\author{Jacek Biero{\'n}}
\affiliation{Instytut Fizyki imienia~Mariana~Smoluchowskiego,
             Uniwersytet Jagiello{\'n}ski \\
             Reymonta~4, 30-059~Krak{\'o}w, Poland \\
             \email{Bieron@uj.edu.pl}}
\author{Charlotte {Froese Fischer}} 
\affiliation{National Institute of Standards and Technology \\
             Gaithersburg, MD 20899-8420, USA}
\author{Paul Indelicato}
\affiliation{Laboratoire Kastler Brossel,
\'Ecole Normale Sup\' erieure \\
 CNRS; Universit\' e P.~et M.~Curie - Paris 6 \\
Case 74; 4, place Jussieu, 75252 Paris CEDEX 05, France}
\author{Per J{\"o}nsson}
\affiliation{Nature, Environment, Society \\
             Malm\"o University, S-205~06 Malm{\"o}, Sweden}
\author{Pekka Pyykk{\"o}}
\affiliation{Department of Chemistry, University of Helsinki \\
             POB 55 (A.I.~Virtasen aukio 1), 00014 Helsinki, Finland}

\date{\today}

\begin{abstract}
The multiconfiguration Dirac-Hartree-Fock (MCDHF) model has been employed
to calculate the expectation values
                for the hyperfine splittings of the
$ 5d^9    6s^2 $~$ ^2 D _{3/2} $ and
$ 5d^9    6s^2 $~$ ^2 D _{5/2} $
levels of atomic gold.
One-, two-, and three-body electron correlation effects
involving all 79 electrons have been included in a systematic manner.
    The approximation employed in this study is             equivalent to
a Complete Active Space (CAS) approach.
Calculated electric field gradients, together with experimental
values of the electric quadrupole hyperfine structure constants,
allow us to extract a nuclear electric quadrupole moment
$Q$($^{197}$Au)=521.5(5.0)~mb.
\end{abstract}


\maketitle


\section{Introduction}
\label{introduction}

{\it Ab initio} calculations of atomic properties
can now be performed routinely, both in the framework of the MCDHF
theory~\cite{GRASP2K,mcdfgme,desclaux:1975,desclaux:1993,GrantBook2007},
as well as many-body perturbation 
(MBPT)
theory~\cite{LindgrenMorrison:86,Dzuba:CI+MBPT:2005,DzubaFlambaum:2007,%
JohnsonBook:2007}.
Both these methods are designed to evaluate
in a systematic manner
the electron-electron correlation effects,
which constitute the dominant correction to all {\it ab initio} calculations
based on the 
central-field 
approach.
%
However, the complexity increases rapidly with the
atomic number, and fully correlated calculations, in which all
electrons are explicitly correlated, are still possible only for
very light elements 
(see~e.g.~\cite{Bieron:Li:1996,Bieron:Be+F:1999,Yerokhin:2008} for
model calculations of hyperfine constants of lithium-like systems).
For heavy atoms both theories can only be applied in a limited
model
(one- and two-body correlation effects)
or only to certain atoms (closed-shell systems or alkali-like systems).
%
The main purpose of the present paper was to carry out an accurate
calculation of hyperfine structure constants of a heavy atom
within the framework of the 
MCDHF theory.
The calculations described in the present paper are, to our knowledge,
the first successful evaluation of one-, two-, and three-body
electron correlation effects for a heavy, open-shell, neutral atom.
The multiconfiguration model applied in the present paper is effectively
equivalent to a Complete Active Space (CAS) approach, in the sense that
in the calculation of the hyperfine electric quadrupole moments
all non-negligible electron correlation effects were explicitly accounted
for at 1~\% level of accuracy or better.
%
The gold atom has been chosen, because the hyperfine
structures~\cite{Autschbach:2002,Malkin:2004,Malkin:2006,Song:2007},
the nuclear electric quadrupole
moments~\cite{Schwerdtfeger:2005,Itano:2006,Yakobi:2007,Belpassi:2007,%
ThierfelderSchwerdtfegerSaue:2007},
and other
properties~\cite{Pyykko:Au:2004,Pyykko:Au:2005,Pyykko:Au:2008}
of gold
have been the subject of much activity recently
(the latest summary of nuclear quadrupole moments is given in                 
   ref.~\cite{Pyykko:08bb}).
The second objective of the present paper  is an evaluation of the
electric quadrupole
 moment $Q$ of the  $ ^{197} $Au  isotope.

\section{Theory}
\label{theory}


The numerical-grid wavefunctions~\cite{GRASP2K}
were generated as the self-consistent solutions of the
Dirac-Hartree-Fock equations~\cite{Grant:1994} in systematically increasing
multiconfiguration bases
(of size
NCF, which is a commonly used shorthand of 'Number of Configuration Functions')
of symmetry-adapted
configuration state functions $ \Phi(\gamma_{k}J) $
\begin{equation}
\label{ASF}
\Psi(J) = \sum_{k}^{NCF} c_{k} \Phi(\gamma_{k}J),
\end{equation}
where $ \Psi(J) $
is an eigenfunction of even parity  and of
total angular momentum $J$
for each of the two states
$ \Psi ( 5d^{9} 6s^{2} \,\, ^2 \! D_{3/2} ) $
and
$ \Psi ( 5d^{9} 6s^{2} \,\, ^2 \! D_{5/2} ) $
of the isotope
$^{197} _{\phantom{1}79}$Au.
The sets 
$ \gamma_{k} $
describe multiconfiguration expansions, for which
configuration mixing coefficients $ c_{k} $ were obtained
through diagonalisation of the
Dirac-Coulomb Hamiltonian
\begin{equation}
\label{Dirac-Coulomb-Hamiltonian}
H_{DC} = \sum_{i} \left[ c {\bm{ \alpha }}_i \cdot
                    {\bm{ p }}_i
         + (\beta_i -1)c^2 + V(r_i) \right]
         + \sum_{i>j} 1/r_{ij}.
\end{equation}
All calculations were done with
the nucleus modelled as a sphere,
where a two-parameter Fermi distribution~\cite{grasp89}
was employed to approximate the radial dependence
of the nuclear charge density.
The nuclear magnetic dipole moment
$\mu$~=~0.145746(9)~$\mu_N$
of $^{197} _{\phantom{1}79}$Au
has been used in calculations of magnetic dipole hyperfine 
constants~\cite{Raghavan:1989,Stone:2005}.

\section{Method}
\label{method}

The numerical wave~functions were obtained independently for the two
levels of interest,
$ 5d^9    6s^2 $~$ ^2 D _{3/2} $ and
$ 5d^9    6s^2 $~$ ^2 D _{5/2} $.
The calculations proceeded in eight phases:
\begin{enumerate}
\item
Spectroscopic orbitals were obtained in the Dirac-Hartree-Fock approximation.
These were kept frozen in all subsequent calculations.
\item
Virtual orbitals were generated in an approximation
(called SrD, and explained in the following subsection 
\ref{virtuals}),
in which all single and restricted double
substitutions from $3spd4spdf5spd6s$ spectroscopic orbitals
to eight layers of virtual orbitals
were included
(see the following subsection 
 \ref{virtuals}
for definitions of spectroscopic
 and virtual orbital sets).
\item
Contributions from $1s2sp$ shells were added in the configuration-interaction
 calculation, i.e.,~with all orbitals frozen.
Only single substitutions contributed to the expectation values.
The configurations involving $1s2sp$ orbitals
 were carried over to the following phases.
\item
%
Unrestricted single and double substitutions (SD) were performed,       
in which one or two occupied orbitals from the $5spd6s$ subshells               
were replaced by orbitals from the virtual set '3spdf2g1h', i.e.,
%
3 virtual orbitals of each of the 's,p,d,f' symmetries, plus
2 virtual orbitals of the 'g' symmetry, and
1 virtual orbital of the 'h' symmetry.
\item
%
Unrestricted triple substitutions (T)
 from $5spd6s$ valence and core orbitals to '2spdf1g' virtual set were added.
\item
The final series of configuration-interaction calculations were based on
the multiconfiguration expansions carried over and merged
from all previous phases enumerated above.
\item
Contributions from the Breit interaction were evaluated in
the single-configuration approximation, including the full Breit operator
in the self-consistent-field process.
\item
The values of the nuclear electric quadrupole moment
$ Q( ^{197} {\rm Au} ) $
were obtained from the relation 
$ B(J)  = 2 e Q \left< JJ | T^{(2)} | JJ \right> $,
where the electronic operator $ T^{(2)} $ represents the electric field
gradient
at the nucleus.
Expectation values of hyperfine constants $A$ and
of electric field gradients were
calculated~\cite{Joensson:hfs:CPC:1996} separately for both states,
$ ^2 D _{3/2} $ and
$ ^2 D _{5/2} $.
The experimental values of the hyperfine 
constants $A$ and $B$ were taken
from~\cite{Blachman:1967,ChildsGoodman:1966}.
\end{enumerate}

\subsection{Virtual orbital set}
\label{virtuals}





%
\begin{table}
\caption{Calculated values of $A$ and $Q$ obtained in
several approximations during the process of generation of virtual
orbital set for the $D_{3/2}$ state.
DHF -- uncorrelated Dirac-Hartree-Fock value;
$n$ -- largest principal quantum number in the orbital set;
$from$ -- spectroscopic orbitals opened for SrD substitutions;
$to$ -- virtual orbital set;
NCF -- number of configurations;
see text for further details.
}
%
\label{calcAQ-D3-2-SrD}
\begin{tabular}{lrlrdd}
\colrule
 \multicolumn{4}{c}{experiment}      & 199.8425(2) &   \\ 
\colrule
 \multicolumn{1}{c}{ n        } &
 \multicolumn{1}{c}{ from } &
 \multicolumn{1}{c}{ to } &
 \multicolumn{1}{r}{ NCF      } &
 \multicolumn{1}{c}{ $A$[MHz] } &
 \multicolumn{1}{c}{ $Q$[mb]  } \\
\colrule  
%
         DHF & \multicolumn{1}{c}{--} & \multicolumn{1}{c}{--} &
                                         1 & 218.011 & 580.807 \\ %
\phantom{1}7 &  5d6s  & 1spdfgh &   1147 & 187.302 & 623.275 \\ %
\phantom{1}8 & 5spd6s & 2spdfgh &  13729 & 198.774 & 652.057 \\ %
\phantom{1}9 & 4spdf...6s & 3spdfgh &  97526 & 195.492 & 547.891 \\ %
%
 10      & 3spd...6s & 4spdfg3h & 222129 & 196.513 & 528.752 \\ 
 11     & 3spd...6s & 5spdfg3h & 222494 & 199.413 & 523.736 \\ 
%
 12      & 3spd...6s & 6spdfg3h & 222851 & 199.455 & 514.186 \\ %
%
 13      & 3spd...6s & 7spdfg3h & 223212 & 200.431 & 515.489 \\ %
%
 14      & 3spd...6s & 8spdfg3h & 223573 & 199.871 & 515.495 \\ 
\colrule  
\end{tabular}
\end{table}
%

%
\begin{table}
\caption{Calculated values of $A$ and $Q$ obtained in
several approximations during the process of generation of virtual
orbital set for the $D_{5/2}$ state.
DHF -- uncorrelated Dirac-Hartree-Fock value;
$n$ -- largest principal quantum number in the orbital set;
$from$ -- spectroscopic orbitals opened for SrD substitutions;
$to$ -- virtual orbital set;
NCF -- number of configurations;
see text for further details.
}
%
\label{calcAQ-D5-2-SrD}
\begin{tabular}{lrlrdd}
\colrule
 \multicolumn{4}{c}{experiment}      & 80.236(3)  &   \\ 
\colrule
 \multicolumn{1}{c}{ n        } &
 \multicolumn{1}{c}{ from   } &
 \multicolumn{1}{c}{ to     } &
 \multicolumn{1}{r}{ NCF      } &
 \multicolumn{1}{c}{ $A$[MHz] } &
 \multicolumn{1}{c}{ $Q$[mb]  } \\
\colrule  
\multicolumn{1}{l}{DHF} &
 \multicolumn{1}{c}{---} &
         \multicolumn{1}{c}{---} &
                                         1 & 79.041 & 612.985 \\ 
\phantom{1}7 &  5d6s  & 1spdfgh &  11984 & 69.487 & 707.216 \\ 
\phantom{1}8 & 5spd6s & 2spdfgh &  33291 & 72.278 & 673.387 \\ 
%
\phantom{1}9 & 4spdf...6s & 3spdfgh & 128639 & 77.761 & 558.526 \\
%
 10 & 3spd...6s  & 4spdfg3h  & 290612 &  81.020 & 532.862 \\ 
%
 11 & 3spd...6s  &  5spdfg3h  & 291039 &  81.045 & 534.635 \\ 
%
 12 & 3spd...6s  &  6spdfg3h  & 291466 &  81.248 & 520.409 \\ 
%
 13 & 3spd...6s  &  7spdfg3h  & 291893 &  81.214 & 520.890 \\ 
%
 14 & 3spd...6s  &  8spdfg3h  & 292320 &  82.136 & 520.259 \\ 
\colrule  
\end{tabular}
\end{table}
%


 
We generated 8 layers of virtual shells
(3 layers with  'spdfgh' symmetries and 5 layers with 'spdfg' symmetries).
It should be noted, that the notion
of a `layer' is somewhat different when applied to occupied 
(also referred to as {\sl spectroscopic}) orbitals,
as opposed to virtual 
(also referred to as {\sl correlation}) orbitals.
A core `layer', i.e.,~a subset
of occupied
orbitals possessing the same 
principal quantum number
(often referred to as a {\sl shell}),
constitutes a set of one-electron spin-orbitals,
clustered in space, and having similar
one-electron energy values.
On the other hand, virtual orbitals with the same principal quantum
number are not necessarily spatially clustered
because
their one-electron energy values do not have physical meaning
and may vary widely, depending on the correlation effects that
a particular virtual orbital describes.
Therefore a `virtual layer' usually
means a subset of the virtual set,
generated in one  step of the procedure,
 as described above.
Such a `layer' is often composed of orbitals with different angular
symmetries.
The notation used in the tables and text of the present paper reflects the 
above considerations, in the sense that occupied orbitals
are listed by their principal and angular quantum numbers
(i.e.~$5spd$ means three occupied orbitals of $s$, $p$, and $d$
 symmetry with principal quantum number $n=5$),
while virtual orbitals are listed by angular symmetry and quantity
(i.e. '5spd' would mean fifteen virtual orbitals ---
 five of each of the 's', 'p', and 'd' symmetries).
To avoid confusion we distinguish occupied orbitals from virtual ones
in the present paper
by using {\sl italics} for occupied orbitals,
while virtual orbitals are enclosed in 'quotation marks'.
This distinction is {\sl not} applied in the tables, since in the tables
there are always headings 'from' and 'to' which clearly denote occupied
and virtual orbitals, respectively.
The notation should always be analysed in the proper context
(see~\cite{Bieron:Au:2008} for further details).
In the present calculations
single and restricted double (SrD) substitutions were allowed
from valence and core orbitals (starting from $5d6s$ for the
first virtual layer).
The restriction was applied to double substitutions in such a way that
only one electron was substituted from core
$3spd4spdf5spd$
shells, the other one had
to be substituted from valence $6s$ shell.
 Each subsequent layer was generated with substitutions
from deeper core shells, down to $3s$.
Table~\ref{calcAQ-D3-2-SrD} shows which occupied orbitals were opened
at each step, as well as composition of the virtual orbital set 
when subsequent layers were generated for the 
$ 5d^9    6s^2 $~$ ^2 D _{3/2} $ state.
For instance, the line marked '10' in the first column describes
the generation of the fourth virtual layer, for which the largest principal
quantum number was 10; all occupied orbitals between $3s$ and $6s$
(i.e.~$3spd4spdf5spd6s$)
were opened for substitutions; the virtual set was composed of 4
orbitals of symmetries 's', 'p', 'd', 'f', 'g', and 3 orbitals
of 'h' symmetry.

The last four layers (those with principal quantum numbers 11, 12, 13, 14)
 were generated with a further restriction,
which allowed only single substitutions to these last layers.

Table~\ref{calcAQ-D5-2-SrD} presents the analogous data obtained
in the process of generation
of virtual orbital layers for the
$ 5d^9    6s^2 $~$ ^2 D _{5/2} $ state.
The data from both tables are
also presented as crosses in figure~\ref{figure2Q2A}.

%
%
%
%

\subsection{Contributions from $1s2sp$ orbitals}
\label{1s2sp}
%

%
%
%

\begin{figure}
%
\resizebox{0.8\columnwidth}{!}{%
  \includegraphics{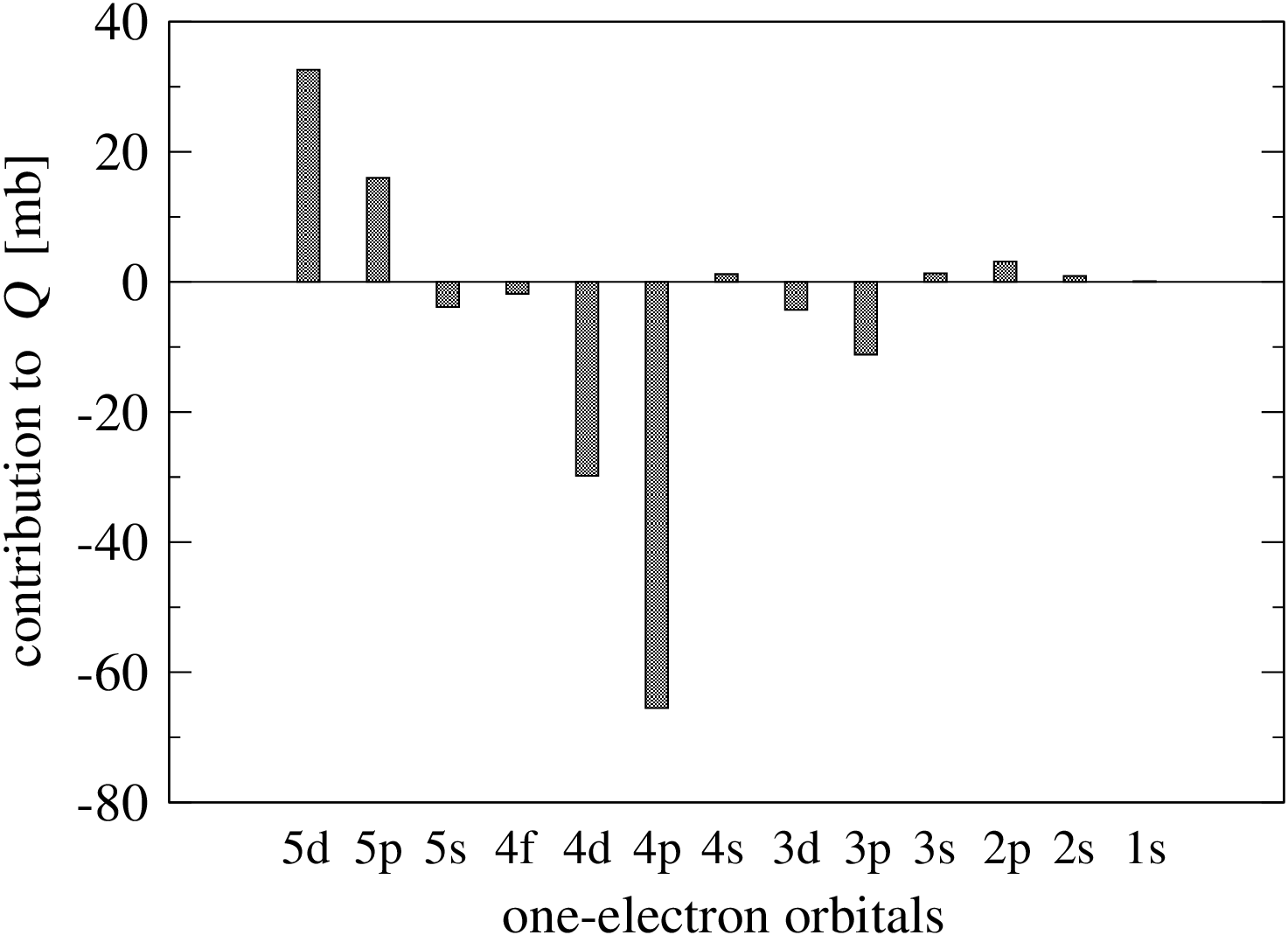} }
\caption{ Contributions from occupied orbitals to the
 calculated value of $Q$ for the 
$ 5d^9    6s^2 $~$ ^2 D _{3/2} $ 
state of Au.
See caption of the Table~\ref{calcAQ-D3-2-1s2sp} and Sec.~\ref{1s2sp}
for further details.
}
\label{figure-1s6s-d3} 
\end{figure}

After generating the virtual orbital set, all orbitals were frozen
and further calculations were carried out in the configuration-interaction
(CI) approach.
First, the effects of $1s2sp$ orbitals were evaluated in separate
CI calculations.
For the 
$ 5d^9    6s^2 $~$ ^2 D _{3/2} $ 
state they are presented in the Table~\ref{calcAQ-D3-2-1s2sp},
together
with the contributions of all other occupied orbitals of the gold atom.
The  orbitals that were open for single and restricted double
 substitutions to the full virtual set are listed in the first column.
The contributions of individual orbitals
(i.e.,~of the leftmost orbital in the first column)
 are listed in the fourth and sixth column and presented in graphical form
in Fig.~\ref{figure-1s6s-d3}.
The individual contributions of the $2p$, $2s$, and $1s$ orbitals
to the total $Q$ value
were of the order of 0.6~\%, 0.2~\%, and 0.02~\%, respectively.
The combined contribution of $1s2sp$ shells
was of the order of 0.8~\%, with respect to the total $Q$ value.
The contribution to the calculated value of magnetic dipole hyperfine
constant $A$ was evaluated  in the same manner as for $Q$.


%
\begin{table}
\caption{ Contributions from occupied orbitals to the
 calculated values of $A$ and $Q$ for the
$ 5d^9    6s^2 $~$ ^2 D _{3/2} $ state of Au;
$orbitals$ = set of orbitals open for single and restricted double
 substitutions from all shells listed in the first column to the full
 virtual set;
NCF = size of the multiconfiguration expansion;
$\Delta$A = contribution [MHz] 
  of the leftmost orbital from a given orbital set to the total $A$ value
(i.e. the individual contribution of the $1s$ orbital
 is listed in the line $1s..6s$);
$\Delta$Q = contribution [mb] 
  of the leftmost orbital from a given set to the $Q$ value.
}
\label{calcAQ-D3-2-1s2sp}
\begin{tabular}{lrdrdr}
%
\colrule
 \multicolumn{1}{r}{ orbitals } &
 \multicolumn{1}{r}{ NCF      } &
 \multicolumn{1}{c}{ $A$[MHz] } &
 \multicolumn{1}{c}{ $\Delta$A } &
 \multicolumn{1}{c}{ $Q$[mb]  } &
 \multicolumn{1}{c}{ $\Delta$Q } \\
\colrule
%
    --- &      1 & 218.011 &
             \multicolumn{1}{c}{ --- } &
                                         580.807 &
                                     \multicolumn{1}{c}{---} \\
           5d6s &  16457 & 189.406 & $-$28.605 & 613.418 &    32.611 \\ %
          5pd6s &  39808 & 153.103 & $-$36.302 & 629.418 &    16.000 \\ %
         5spd6s &  48129 & 190.849 &    37.746 & 625.559 &  $-$3.859 \\ %
       4f5spd6s &  89477 & 187.661 &  $-$3.188 & 623.738 &  $-$1.821 \\ %
      4df5spd6s & 124673 & 194.614 &     6.953 & 593.938 & $-$29.800 \\ %
     4pdf5spd6s & 148188 & 202.721 &     8.107 & 528.442 & $-$65.496 \\ %
    4spdf5spd6s & 156525 & 196.476 &  $-$6.245 & 529.646 &     1.204 \\ %
  3d4spdf5spd6s & 191721 & 199.346 &     2.870 & 525.342 &  $-$4.304 \\ %
       3pd...6s & 215236 & 201.106 &     1.760 & 514.175 & $-$11.167 \\ %
      3spd...6s & 223573 & 199.872 &  $-$1.234 & 515.495 &     1.320 \\ %
    2p3spd...6s & 247088 & 196.564 &  $-$3.308 & 518.635 &     3.140 \\ %
   2sp3spd...6s & 255425 & 199.576 &     3.012 & 519.539 &     0.904 \\ %
 1s2sp3spd...6s & 263762 & 199.554 &  $-$0.022 & 519.634 &     0.095 \\ %
\colrule
%
%
%
\end{tabular}
\end{table}
%

\begin{figure}
%
\resizebox{0.8\columnwidth}{!}{%
  \includegraphics{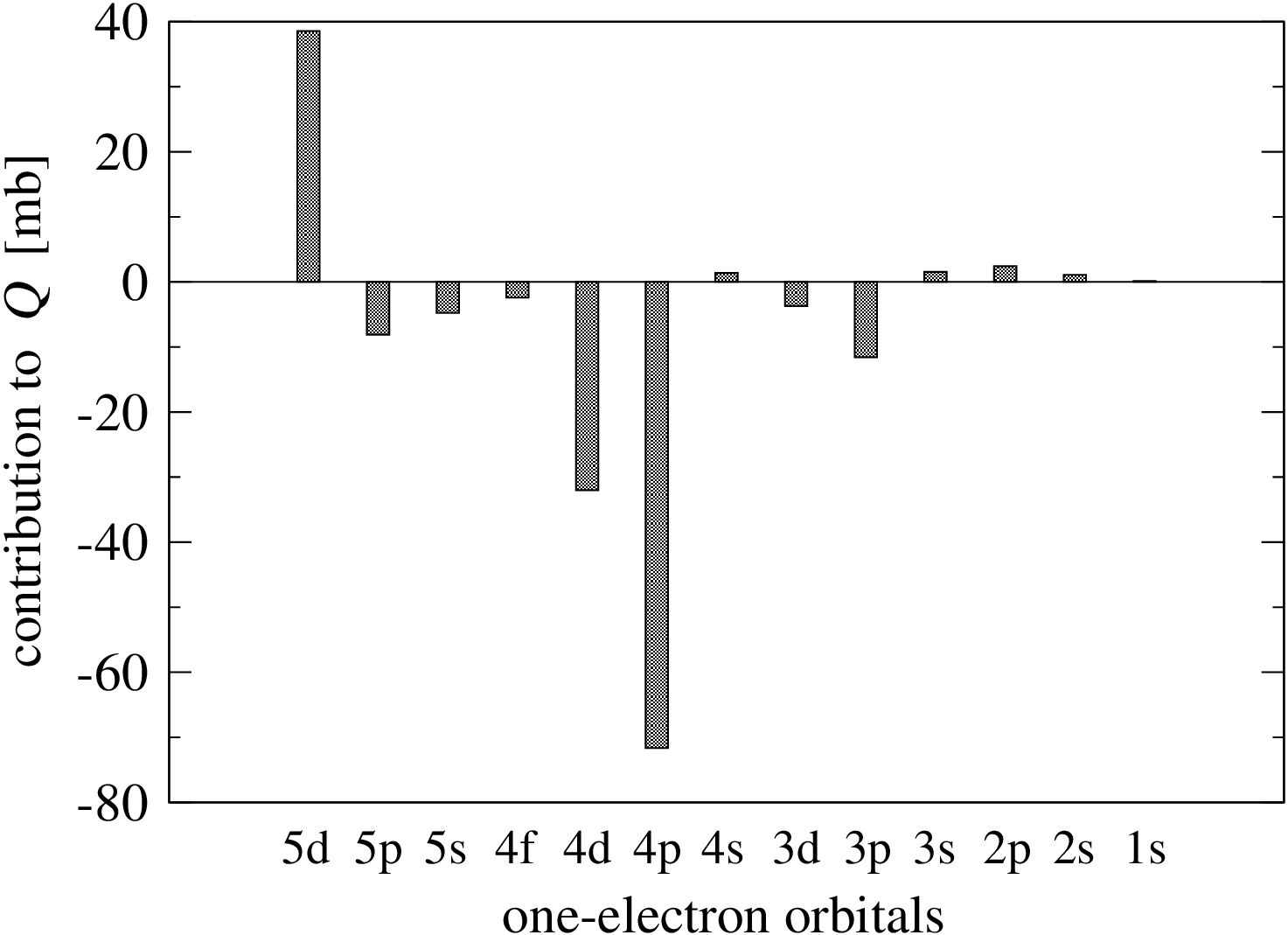} }
\caption{ Contributions from occupied orbitals to the
 calculated value of $Q$ for the 
$ 5d^9    6s^2 $~$ ^2 D _{5/2} $ 
state of Au.
See caption of Table~\ref{calcAQ-D5-2-1s2sp} and Sec.~\ref{1s2sp}
for further details.
}
\label{figure-1s6s-d5} 
\end{figure}

A similar procedure has been carried out for the $Q$ and $A$ values
of the
$ 5d^9    6s^2 $~$ ^2 D _{5/2} $ state.
The results for the $ ^2 D _{5/2} $ state are shown in
table~\ref{calcAQ-D5-2-1s2sp}
and in
figure~\ref{figure-1s6s-d5}.
The individual contributions of the $2p$, $2s$, and $1s$ orbitals
to the total $Q$ value
were of the order of 0.5~\%, 0.2~\%, and 0.02~\%, respectively.
The combined contribution of $1s2sp$ shells
was of the order of 0.7~\%, with respect to the total $Q$ value.

All these contributions have been
included in the $Q$ and $A$ values obtained within the SrD approximation
%
and the configuration state functions (CSFs) involved in evaluation
of these contributions were carried over to all subsequent calculations.

It should be pointed out that
the data in Tables~\ref{calcAQ-D3-2-1s2sp} and~\ref{calcAQ-D5-2-1s2sp}
and in the Figures~\ref{figure-1s6s-d3} and~\ref{figure-1s6s-d5}
were obtained with single and restricted double
substitutions, i.e.,~with unrestricted double and triple
substitutions excluded.
Therefore the contributions of the $5psd$ and $4spdf$ shells
are somewhat distorted ---
 if double and triple
  substitutions were included,
  the individual contributions of the $5psd$ and $4spdf$ shells
  would differ by a few percent.
Only the $3spd$, $2sp$ and $1s$
shells are essentially insensitive to double and triple
substitutions (see Sec.~\ref{sdtq} below).
Therefore their contributions are approximately correct.

%
\begin{table}
\caption{ Contributions from occupied orbitals to the
 calculated values of $A$ and $Q$ for the 
$ 5d^9    6s^2 $~$ ^2 D _{5/2} $ state of Au;
$orbitals$ = set of orbitals open for single and restricted double
 substitutions from all shells listed in the first column to the full
 virtual set;
NCF = size of the multiconfiguration expansion;
$\Delta$A = contribution [MHz] 
  of the leftmost orbital from a given orbital set to the total $A$ value
(i.e. the individual contribution of the $1s$ orbital
 is listed in the line $1s..6s$);
$\Delta$Q = contribution [mb] 
  of the leftmost orbital from a given set to the $Q$ value.
%
%
}
\label{calcAQ-D5-2-1s2sp}
\begin{tabular}{rrdrdr}
\colrule
 \multicolumn{1}{r}{ orbitals } &
 \multicolumn{1}{r}{ NCF      } &
 \multicolumn{1}{c}{ $A$[MHz] } &
 \multicolumn{1}{c}{ $\Delta$A } &
 \multicolumn{1}{c}{ $Q$[mb]  } &
 \multicolumn{1}{c}{ $\Delta$Q } \\
\colrule  
%
    --- &      1 & 79.041 &
             \multicolumn{1}{c}{ --- } &
                                        612.985 &
                                     \multicolumn{1}{c}{---} \\
          5d6s  &  21501 & 106.724 &    27.683 & 651.547 &    38.562 \\ %
         5pd6s  &  51800 & 109.554 &     2.830 & 643.451 &  $-$8.096 \\ %
        5spd6s  &  62536 &  71.472 & $-$38.082 & 638.694 &  $-$4.757 \\ %
      4f5spd6s  & 117626 &  70.636 &  $-$0.836 & 636.280 &  $-$2.414 \\ %
     4df5spd6s  & 163739 &  73.490 &     2.854 & 604.280 & $-$32.000 \\ %
    4pdf5spd6s  & 194221 &  74.597 &     1.107 & 532.632 & $-$71.648 \\ %
   4spdf5spd6s  & 204973 &  79.767 &     5.170 & 534.008 &     1.376 \\ %
 3d4spdf5spd6s  & 251086 &  80.877 &     1.110 & 530.308 &  $-$3.700 \\ %
      3pd...6s  & 281568 &  80.580 &  $-$0.297 & 518.734 & $-$11.574 \\ %
     3spd...6s  & 292320 &  82.136 &     1.556 & 520.259 &     1.525 \\ %
   2p3spd...6s  & 322802 &  81.700 &  $-$0.436 & 522.677 &     2.418 \\ %
  2sp3spd...6s  & 333554 &  78.995 &  $-$2.705 & 523.757 &     1.080 \\ %
1s2sp3spd...6s  & 344306 &  79.025 &     0.030 & 523.880 &     0.123 \\ %
\colrule
\end{tabular}
\end{table}
%

\subsection{Double, triple, and quadruple substitutions}
\label{sdtq}

The decomposition of the electron
correlation correction to the hyperfine structure
into one-, two-, three-, and four-body effects can be understood
from the following (simplified) analysis.
The structure of the
$ 5d^9 6s^2 $~$ ^2 D $ 
states of gold is determined
to a large extent by the interaction of the valence $6s^2$ shell with
a highly polarisable $5d^{9}$ shell.
The direct and indirect effects of relativity
bring the outer $d$ shell much closer,
radially and energetically,
to the valence $s$
orbital than in homologous silver and copper
atoms~\cite{Pyykko:1988,Pyykko:Au:2004}.
This in turn increases the polarisation of the $5d^{9}$ shell
by the valence electrons.
Therefore,
the core-valence interaction
(the leading electron correlation correction)
 leads to the contraction of
the $6s$ orbital, which overestimates the hyperfine structure.
The unrestricted double substitutions affect the hyperfine structure in
 two ways: directly through the 
configuration state functions (CSFs) 
themselves but
 also indirectly through the change
of the expansion coefficients of the important configurations
 obtained by single substitutions.
%
Three-particle effects in turn affect the expansion coefficients
of the configurations obtained from double substitutions.
In a simple picture we can describe the wave function in terms
 of pair-correlation functions and the three-particle effects then account
 for polarisation of pair-correlation functions,
 leading to an increase of the hyperfine
 structure~\cite{Joensson:hfsNa:1996}.
Four-particle effects
affect mostly the expansion coefficients
of the configurations obtained from double substitutions.
Therefore their influence on the hyperfine structure is indirect
and second-order to that of the double substitutions.
They are usually small and can often be
 neglected~\cite{Godefroid:1997};
 they are discussed in Sec.~\ref{sdtq}.

Tables~\ref{calcAQ-D3-2-SD} and~\ref{calcAQ-D5-2-SD}
show the results of configuration-interaction calculations,
where
various combinations of occupied and virtual sets were
tested with single and unrestricted double substitutions.
The data from both tables are
also presented as empty circles in Fig.~\ref{figure2Q2A}. 
%
%
%

%
\begin{table}
\caption{Values of $Q$ and $A$ for the $D_{3/2}$ state,
calculated in configuration-interaction approach, with
single and unrestricted double substitutions, in
several different multiconfiguration expansions;
$from$ -- spectroscopic orbitals opened for substitutions;
$to$ -- virtual orbital set;
NCF -- number of configurations;
see text for further details.
%
%
}
\label{calcAQ-D3-2-SD}
\begin{tabular}{llrdd}
\colrule  
 \multicolumn{3}{c}{experiment}      & 199.8425(2) &   \\ 
\colrule
 \multicolumn{1}{c}{ from     } &
 \multicolumn{1}{c}{ to       } &
 \multicolumn{1}{c}{ NCF      } &
 \multicolumn{1}{c}{ $A$[MHz] } &
 \multicolumn{1}{c}{ $Q$[mb]  } \\
\colrule  
%
 5spd6s & 1spdfgh & 259135 & 205.426 & 521.191 \\ 
 4spdf5spd6s & 1spdfgh & 358019 & 205.968 & 521.503 \\ 
 5spd6s & 2spdf   & 279559 & 210.523 & 509.839 \\ 
 5spd6s &2spdfg  & 320545 & 211.512 & 511.461 \\ 
 5spd6s & 2spdfgh & 366257 & 211.480 & 512.286 \\ 
 5spd6s & 3spdf2g1h &459594 & 213.088 & 510.451 \\ 
 5spd6s & 3spdf2gh & 465794 & 213.075 & 510.402 \\ 
 5spd6s & 3spdfg2h & 506987 & 213.146 & 510.268 \\ 
 5spd6s & 4spdf2gh & 687301 & 213.200 & 510.478 \\ 
\colrule  
%
%
%
\end{tabular}
\end{table}
%

%
\begin{table}
\caption{Values of $Q$ and $A$ for the $D_{5/2}$ state,
calculated in configuration-interaction approach, with
single and unrestricted double substitutions, in
several different multiconfiguration expansions;
$from$ -- spectroscopic orbitals opened for substitutions;
$to$ -- virtual orbital set;
NCF -- number of configurations;
see text for further details.
%
%
}
\label{calcAQ-D5-2-SD}
\begin{tabular}{rlrdd}
\colrule
 \multicolumn{3}{c}{experiment}      & 80.236(3)  &   \\ 
\colrule
 \multicolumn{1}{c}{ from     } &
 \multicolumn{1}{c}{ to       } &
 \multicolumn{1}{c}{ NCF      } &
 \multicolumn{1}{c}{ $A$[MHz] } &
 \multicolumn{1}{c}{ $Q$[mb]  } \\
\colrule  
%
 5spd6s & 1spdfgh & 339306 &  74.258 & 507.823 \\ 
 4spdf5spd6s & 1spdfgh & 467381 &  72.048 & 509.321 \\ 
 5spd6s & 2spdfgh & 480824 &  73.468 & 512.278 \\ 
 5spd6s & 3spdf2gh & 607421 &  73.494 & 512.559 \\ 
 5spd6s & 4spdf3gh & 898368 &  73.294 & 514.621 \\ 
 5spd6s & 5spdf4g3h  & 1228675 &  73.212 & 514.269 \\ 
\colrule  
\end{tabular}
\end{table}
%

%
\begin{table}
\caption{Values of $Q$ and $A$ for the $D_{3/2}$ state,
calculated in configuration-interaction approach, with
single and unrestricted double and triple substitutions, in
several different multiconfiguration expansions;
$from$ -- spectroscopic orbitals opened for substitutions;
$to$ -- virtual orbital set;
NCF -- number of configurations;
see text for further details.
%
%
%
}
\label{calcAQ-D3-2-SDT}
\begin{tabular}{rlrdd}
\colrule
 \multicolumn{3}{c}{experiment}      & 199.8425(2) &   \\ 
\colrule
 \multicolumn{1}{c}{ from     } &
 \multicolumn{1}{c}{ to       } &
 \multicolumn{1}{c}{ NCF      } &
 \multicolumn{1}{c}{ $A$[MHz] } &
 \multicolumn{1}{c}{ $Q$[mb]  } \\
%
\colrule
%
 5spd6s &   1spd           &  265183 & 198.955 & 520.346 \\ 
 5spd6s &   1spdf          &  386326 & 194.391 & 533.464 \\ 
 5spd6s &   1spdfg         &  641227 & 193.744 & 536.620 \\ 
 5spd6s &   1spdfgh        & 1012615 & 193.246 & 537.786 \\ 
%
 5spd6s &    2spd1f        &  943544 & 198.752 & 522.361 \\ 
%
 5spd6s & 2spdf            & 1543051 & 199.973 & 520.536 \\ 
%
 5spd6s & 3spd2f          & 1200261 & 198.207 & 520.267 \\ 
 5spd6s & 3psdf           & 1309130 & 198.254 & 520.096 \\ 
\colrule
\end{tabular}
\end{table}
%

%
\begin{table}
\caption{Values of $Q$ and $A$ for the $D_{5/2}$ state,
calculated in configuration-interaction approach, with
single and unrestricted double and triple substitutions, in
several different multiconfiguration expansions;
$from$ -- spectroscopic orbitals opened for substitutions;
$to$ -- virtual orbital set;
NCF -- number of configurations;
see text for further details.
%
%
%
}
\label{calcAQ-D5-2-SDT}
\begin{tabular}{rlrdd}
\colrule
 \multicolumn{3}{c}{experiment}      &  80.236~(3) &   \\ 
\colrule
 \multicolumn{1}{c}{ from     } &
 \multicolumn{1}{c}{ to       } &
 \multicolumn{1}{c}{ NCF      } &
 \multicolumn{1}{c}{ $A$[MHz] } &
 \multicolumn{1}{c}{ $Q$[mb]  } \\
\colrule  
%
 5spd6s & 1spd           &  341440 & 81.6955 & 514.929 \\ 
 5spd6s & 1spdf          &  456506 & 82.1357 & 520.259 \\ 
 4f5spd6s &   1spdf        & 1403860 & 79.9343 & 518.380 \\ 
%
 5spd6s & 1spdfg         &  842883 & 80.2371 & 519.291 \\ 
%
 5spd6s & 2spdf          & 1326851 & 83.0623 & 521.862 \\ 
\colrule
%
\end{tabular}
\end{table}

%
\begin{table}
\caption{Values of $Q$ and $A$ for the $D_{3/2}$ state,
calculated in configuration-interaction approach, with
single and unrestricted double, triple, and quadruple substitutions,
in several different multiconfiguration expansions;
$type$ -- substitution multiplicity;
$from$ -- spectroscopic orbitals opened for substitutions;
$to$ -- virtual orbital set;
NCF -- number of configurations;
see text for further details.
%
%
}
\label{calcAQ-D3-2-SDTQ}
\begin{tabular}{lrlrdd}
\colrule
 \multicolumn{3}{c}{experiment}      & 199.8425(2) &   \\ 
\colrule
 \multicolumn{1}{c}{ type     } &
 \multicolumn{1}{c}{ from     } &
 \multicolumn{1}{c}{ to       } &
 \multicolumn{1}{c}{ NCF      } &
 \multicolumn{1}{c}{ $A$[MHz] } &
 \multicolumn{1}{c}{ $Q$[mb]  } \\
\colrule  
%
SDT  & 5spd6s & 1spd     & 386326 & 194.391 & 533.464 \\ 
SDTQ & 5spd6s & 1spd     & 569497 & 194.301 & 533.653 \\%
SDT  & 5spd6s & 1spdf    &  386326 & 194.391 & 533.464 \\ 
SDTQ & 5spd6s & 1spdf    & 967871 & 195.376 & 531.685 \\%
SDTQ & 5spd6s & 1spdf    & 1089014 & 194.686 & 531.846 \\
\colrule
\end{tabular}
\end{table}
%

The second line in
Tables~\ref{calcAQ-D3-2-SD} and~\ref{calcAQ-D5-2-SD}
represents a calculation in which substitutions
from the $4spdf$ shells were allowed to one layer of virtual orbitals.
When compared with the first line, it yields the effect of
$4spdf$ shells on the calculated values of $Q$ and $A$.
In order to limit the size of the configuration expansions,
the CSFs representing the above substitutions were not carried over
to the following, higher order calculations.
Instead, the corrections were
included additively, as described in Sec.~\ref{corrections}.
At the same time the evaluation of these corrections may be
treated as a crude estimate of error arising from omitted
double substitutions from occupied shells
(see Sec.~\ref{error} for details).

An inspection of the last column of 
Table~\ref{calcAQ-D3-2-SD}
indicates, that three layers of virtual orbitals were necessary
to reach convergence of the $Q$ and $A$ values in the
{\sl single and unrestricted double substitutions} (SD)
approximation for the 
$ 5d^9    6s^2 $~$ ^2 D _{3/2} $ state.
Four layers were necessary in case of the
$ 5d^9    6s^2 $~$ ^2 D _{5/2} $ state
(see Table~\ref{calcAQ-D5-2-SD}).

Tables~\ref{calcAQ-D3-2-SDT} and~\ref{calcAQ-D5-2-SDT}
show the results of configuration-interaction calculations,
in which various combinations of occupied and virtual sets were
tested with unrestricted double and triple substitutions.
The data from both tables are
also presented as full circles in Fig.~\ref{figure2Q2A}. 
Two layers of virtual orbitals were necessary
to reach convergence of the $Q$ value in the
{\sl single and double and triple substitutions} (SDT)
approximation for both
$ 5d^9    6s^2 $~$ ^2 D _{3/2} $ and
$ 5d^9    6s^2 $~$ ^2 D _{5/2} $ states.
In case of the $A$ values, convergence required
three, rather than two, layers.

Table~\ref{calcAQ-D3-2-SDTQ} shows the effect of quadruple substitutions.
The first line represents an approximation in which
single, double, and triple substitutions from $5spd6s$ orbitals
to a truncated virtual layer composed of 's', 'p', and 'd' symmetries
were included.
The third line represents a similar approximation in which the
(still truncated) virtual layer was composed of 's', 'p', 'd', and 'f'
symmetries.
The second, fourth, and fifth lines represent corresponding
'quadruple approximation'
in which single, double, triple, and quadruple substitutions were
allowed.
The comparison had to be made on a
reasonably small orbital set in order to be able to converge
the calculation involving quadruple substitutions.
The numbers of CSFs in the last two lines are different, because
certain restrictions were applied in the calculation represented
by the fourth line (see the comments near the end of 
Sec.~\ref{configuration-interaction} for details).
The results presented in the table~\ref{calcAQ-D3-2-SDTQ} indicate
that the correction involving quadruple substitutions
is unlikely to exceed 1~\%.
The CSFs representing quadruple substitutions were not carried over
to the following calculations and the 'quadruple' correction was
included additively, as described in Sec.~\ref{corrections}.

\subsection{4-D configuration-interaction calculations}
\label{configuration-interaction}

A full, converged Complete Active Space (CAS) calculation
for the gold atom is still unattainable due to software and hardware
limitations.
Based on our current calculations
we estimate that the CAS approach would require 
configuration expansions in four 'dimensions':
(1) single, double, triple, and perhaps quadruple substitutions;
(2) from all core shells (or at least from  $3spd4spdf5spd6s$);
(3) to eight or more virtual orbital layers;
(4) of $s$, $p$, $d$, $f$, $g$, $h$, and perhaps higher symmetries.
One can imagine a 'space'
spanned by the four 'dimensions' defined above,
i.e.~'{\sl substitution} multiplicity',
'number of opened {\sl core} subshells', 
'number of {\sl virtual} layers',
and
'maximal {\sl symmetry} of virtual layers' dimension.
In fact, this space should rather be called a 'matrix',
 since all four dimensions are discrete. Let's call
this four-dimensional matrix a 'CAS matrix'.
Each element of the matrix is represented by a multiconfiguration
expansion obtained by substituting
a particular number of electrons ('substitution' dimension)
from specific core orbitals ('core' dimension)
to a set of virtual orbitals ('virtual' dimension)
of specific symmetries ('symmetry' dimension).
A full CAS calculation would require several orders of magnitude larger
configuration expansions than are possible even with
the largest computer resources available today.

%
\begin{table}
\caption{The final configuration-interaction calculations
 of  $Q$ and $A$ for the
 $ 5d^9    6s^2 $~$ ^2 D _{3/2} $ state of Au;
$type$ --- description of the multiconfiguration expansions,
 see text for details;
NCF = size of the multiconfiguration expansion.
%
%
}
\label{calcAQ-D3-2-F}
\begin{tabular}{lrdd}
\colrule
 \multicolumn{2}{c}{experiment}      & 199.8425(2) &   \\ 
\colrule
 \multicolumn{1}{c}{ type     } &
 \multicolumn{1}{c}{ NCF      } &
 \multicolumn{1}{c}{ $A$[MHz] } &
 \multicolumn{1}{c}{ $Q$[mb]  } \\
\colrule  
%
%
%
  SD:3hgg \!\!\! + SDT:2fd  & 1182329 & 206.343 & 517.201 \\ %
  SD:3hgf        + SDT:2fd  & 1144532 & 206.221 & 517.342 \\ %
%
  SD:3hgf        + SDT:2gd  & 1711382 & 205.104 & 519.106 \\ %
  SD:3hgf        + SDT:2gf  & 1847380 & 204.489 & 519.829 \\ %
%
\colrule
\end{tabular}
\end{table}
%

%
\begin{table}
\caption{The final configuration-interaction calculations
 of  $Q$ and $A$ for the
 $ 5d^9    6s^2 $~$ ^2 D _{5/2} $ state of Au;
$type$ --- description of the multiconfiguration expansions,
 see text for details;
NCF = size of the multiconfiguration expansion.
%
%
%
}
\label{calcAQ-D5-2-F}
\begin{tabular}{lrdd}
\colrule
 \multicolumn{2}{c}{experiment}      &  80.236~(3) &   \\ 
\colrule
 \multicolumn{1}{c}{ type     } &
 \multicolumn{1}{c}{ NCF      } &
 \multicolumn{1}{c}{ $A$[MHz] } &
 \multicolumn{1}{c}{ $Q$[mb]  } \\
\colrule  
 SD:3hgf + SDT:2fd  & 1441120 & 78.2451 & 520.073 \\ %
 SD:3hgf + SDT:2gd  & 1527668 & 79.9182 & 522.066 \\ %
%
\colrule
\end{tabular}
\end{table}

However, a computational strategy can be designed in which 
a
considerably smaller
multiconfiguration expansion yields a wave~function only marginally
inferior to a full CAS wave~function, in the sense that all important
electron correlation effects are included
and the calculated values of $A$ and $Q$ are close to those that
would result from a full, converged, CAS calculation.
%
The strategy is based on the observation that one does not have to
simultaneously
push the configuration expansions to the limits of all
the above mentioned 'dimensions'.
Specifically, the dependence of atomic properties on the
 'substitution' dimension is critical.
To illustrate this approach, let's consider separately
the contributions of single, double, and triple substitutions
to the calculated values of $A$ and $Q$ of gold.
To obtain converged results within a {\sl single substitutions} model,
one has to include substitutions from all occupied shells
($1s2sp3spd4spdf5spd6s$)
to eight or more virtual layers.
This is illustrated in
tables~\ref{calcAQ-D3-2-SrD},
\ref{calcAQ-D5-2-SrD},
\ref{calcAQ-D3-2-1s2sp}, and
\ref{calcAQ-D5-2-1s2sp},
where eight
virtual orbital layers were necessary to converge the
series of self-consistent-field calculations.
However, to obtain a converged result within a
{\sl single and double substitutions} model (SD)
one has to include double substitutions from 
$4spdf5spd6s$ occupied orbitals, not to eight, but
to three or at most four virtual layers (see 
Tables~\ref{calcAQ-D3-2-SD} and~\ref{calcAQ-D5-2-SD}).
In the {\sl single, double, and triple substitutions} model (SDT)
it is enough to consider triple substitutions from
$5spd6s$ occupied orbitals
to two or at most three virtual layers (see
Tables~\ref{calcAQ-D3-2-SDT} and~\ref{calcAQ-D5-2-SDT}).
In the 'space' (or rather in the 'matrix')
of the four 'dimensions' defined above,
the 'core' and 'virtual' dimension sizes strongly depend on the
'substitution' dimension
 (in fact, all four dimensions are interdependent).

Therefore, one can construct an approximation, in which all important
electron correlation effects are included
and the calculated values of $A$ and $Q$ are close to those that
would result from a full, converged CAS calculation.
In order to find a suitable approximation, we have performed
a set of test calculations for several elements of the above mentioned
'matrix'.
For each 'dimension', the calculations were saturated to the point
where the relative change of the expectation values
(i.e.,~both $A$ and $Q$) did not exceed a small fraction of a percent
(usually two or three tenths of a percent).
Specifically, for each 'substitution' dimension
(i.e.,~for single, double, and triple substitutions)
we thoroughly tested the dependence of observables
on 'symmetry,' 'virtual', and 'core' spaces.
When a saturated set of configuration state functions (CSFs) is obtained for
a particular 'substitution' dimension,
all these CSFs are carried over to the next step(s).
The merged, 'final' multiconfiguration expansion represents an approximation,
which is effectively equivalent to a CAS expansion,
and the corresponding wavefunction is of similar quality as a
CAS wavefunction, at least from the point of view of the
calculated values of $A$ and $Q$.

\begin{figure*}%
     \subfigure[ ~from electric field gradient and
                  B($ 5d^9 6s^2 $~$ ^2 D_{3/2} $) ]{
          \label{subfigureQD3}
          \includegraphics[width=.448\textwidth]{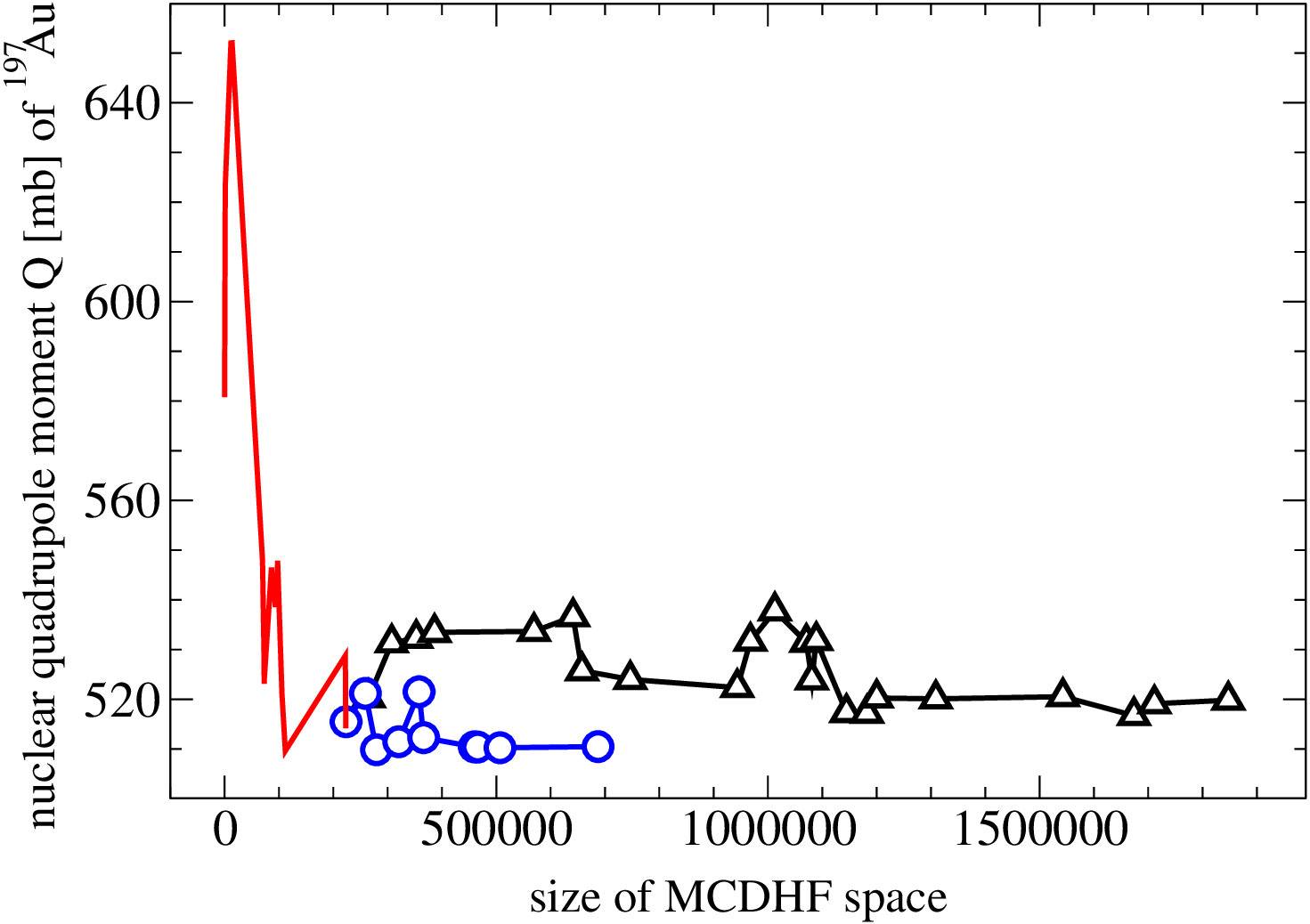}}
     \hspace{.3in}
     \subfigure[ ~from electric field gradient and
                  B($ 5d^9 6s^2 $~$ ^2 D_{5/2} $) ]{
          \label{subfigureQD5}
          \includegraphics[width=.44\textwidth]{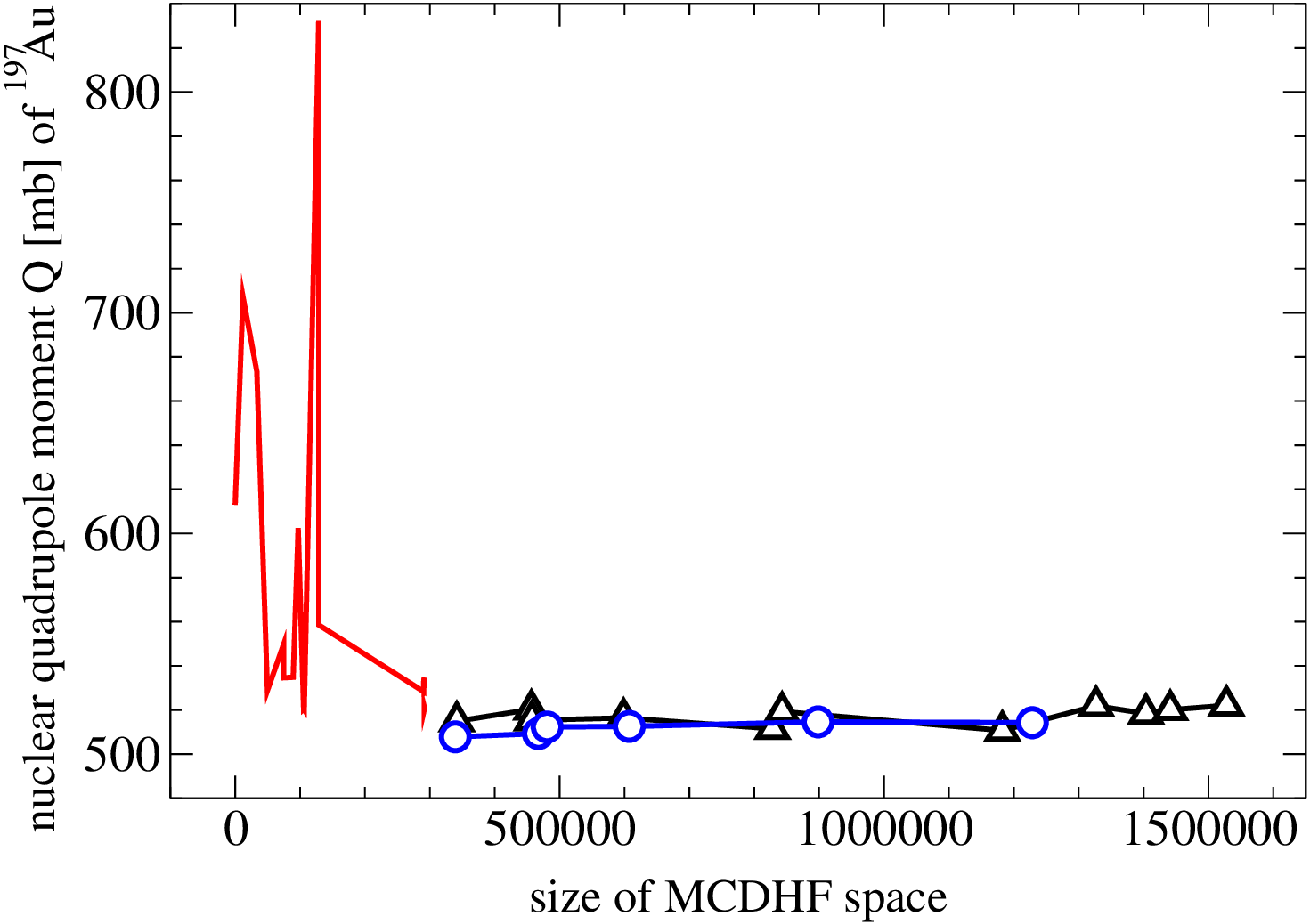}} \\
     \vspace{.3in}
     \subfigure[ ~A($ 5d^9 6s^2 $~$ ^2 D_{3/2} $) ]{
          \label{subfigureAD3}
          \includegraphics[width=.448\textwidth]{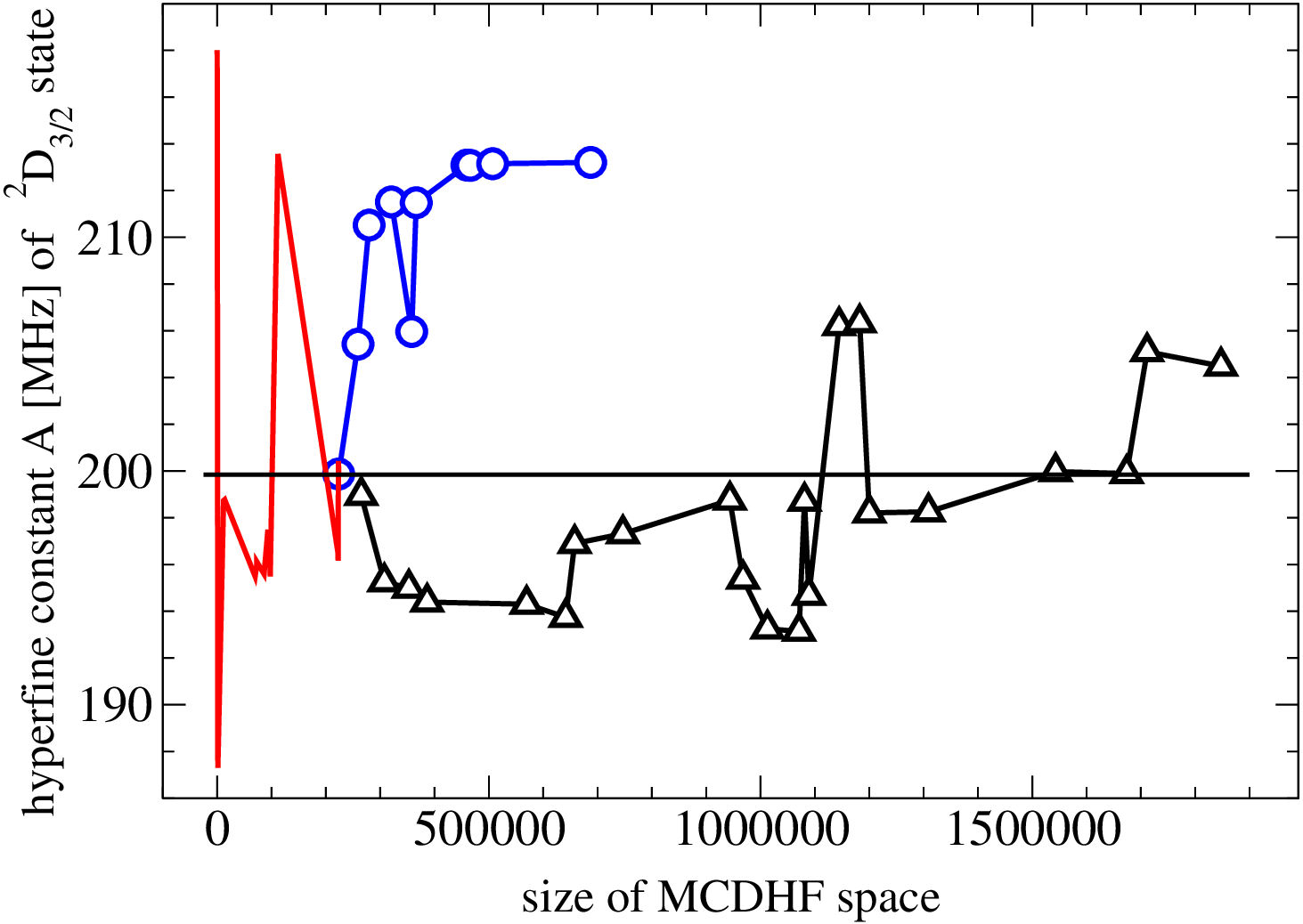}}
     \hspace{.3in}
     \subfigure[ ~A($ 5d^9 6s^2 $~$ ^2 D_{5/2} $) ]{
          \label{subfigureAD5}
          \includegraphics[width=.44\textwidth]{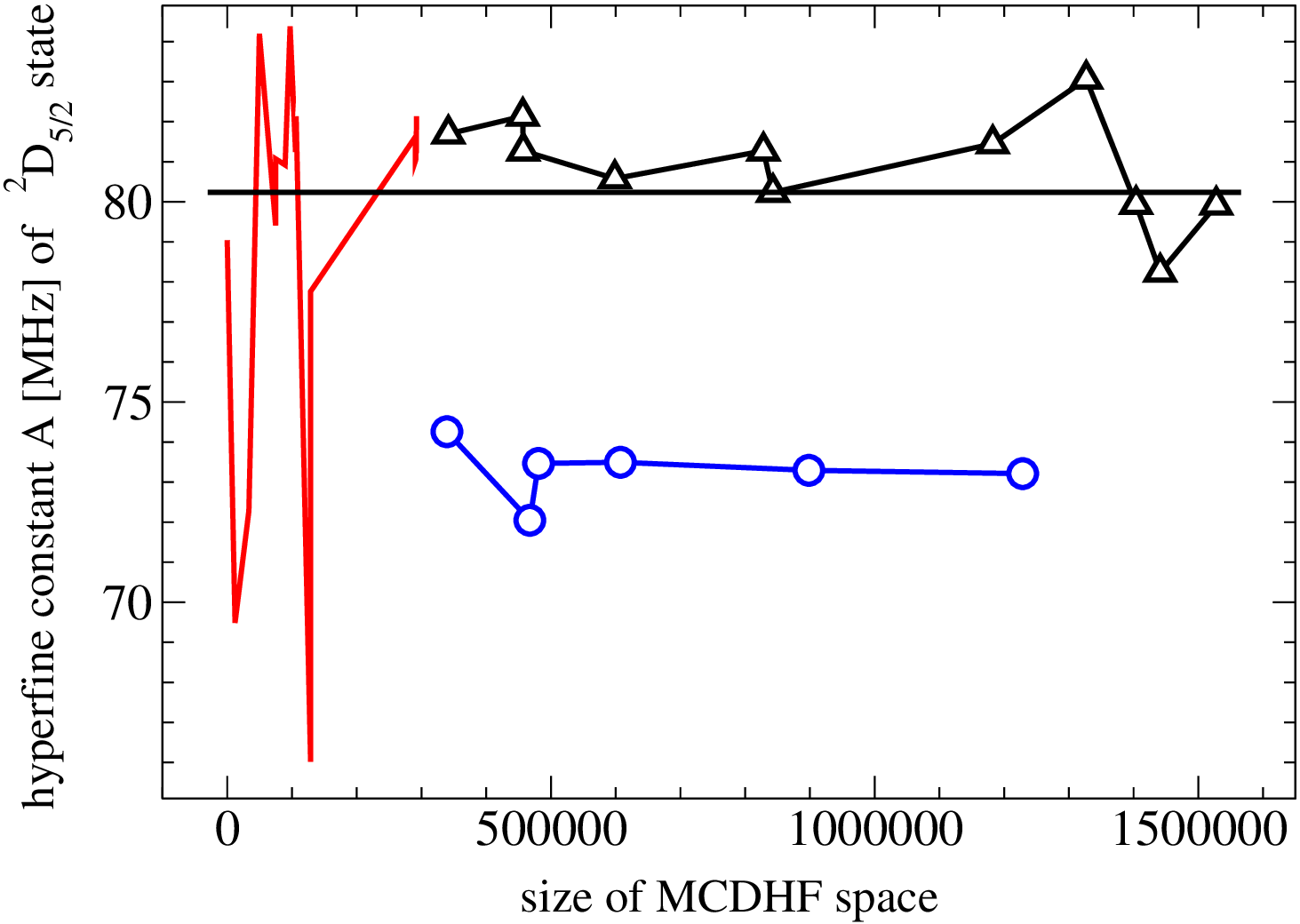}}
     \caption{(Color online)
              Nuclear quadrupole moment $Q$($^{197}\!$Au) [mb] 
 obtained from the calculated electric field gradients,
              and hyperfine magnetic dipole constants $A$ [MHz]
              of the
              states $ 5d^9 6s^2 $~$ ^2 D_{3/2} $ 
              and
                     $ 5d^9 6s^2 $~$ ^2 D_{5/2} $,
  as functions of the size of the multiconfiguration expansions;
line with no symbols (red online) -- 'SrD' approximation;
circles (blue online) -- 'SD' approximation;
triangles (black online) -- 'SDT' and final 'CAS' approximations
(see text for details).
Horizontal straight lines in figures (c) and (d) represent the experimental
values of hyperfine constants
 $ A ( ^2 D_{3/2} ) $~=~199.8425(2)~MHz~\cite{Blachman:1967} 
and 
 $ A ( ^2 D_{5/2} ) $~=~80.236(3)~MHz~\cite{ChildsGoodman:1966},
respectively. 
The small corrections described in subsection~\ref{corrections}
are not included in the figures.
}
\label{figure2Q2A} 
\end{figure*}

In practice there is not one single final 'CAS' expansion,
but a series
of such final expansions in which various sets
of 'S', 'SD', and 'SDT' multiconfiguration expansions
(i.e.~various sets with single, double, and triple substitutions)
are merged together.
Table~\ref{calcAQ-D3-2-F}
shows the results obtained from
a series of such final 'CAS' calculations for the $ ^2 D _{3/2} $ state,
and 
Table~\ref{calcAQ-D5-2-F} shows the same for the $ ^2 D _{5/2} $ state.
The data from both tables are
also included in Fig.~\ref{figure2Q2A}.
%
%
The 'CAS' expansions are composed as follows.
All virtual orbitals and all CSFs
generated in the SrD approximation,
as described in section~\ref{virtuals},
as well as those described in section~\ref{1s2sp},
were included.
The remaining CSF expansions were generated with substitutions
from $5spd6s$ orbitals to virtual sets described in the first
column of Tables~\ref{calcAQ-D3-2-F} and~\ref{calcAQ-D5-2-F},
where
symbols before the colon represent substitution multiplicity, i.e.,
SD --- single and double substitutions;
SDT --- single and double and triple substitutions;
the symbols after the colon represent virtual orbital layers, i.e.,
3hgg --- three layers
(first layer with 'spdfgh' symmetries and two layers with 'spdfg' symmetries);
2fd --- two layers
(first layer with 'spdf' symmetries and second layer with 'spd' symmetries);
3hgf --- three layers
(first layer with 'spdfgh' symmetries,
 second layer with 'spdfg' symmetries;
 third layer with 'spdf' symmetries), etc.

In the largest calculations, when single, double, and triple substitutions
to two or three layers were included, we had to further  limit the overall
number of CSFs, due to software and hardware limitations.
In those cases,
 the occupation number of the least important virtual
orbital was restricted to single or double, thus excluding those CSFs
in which this particular virtual orbital was occupied by three electrons.
The difference that such a restriction brings about
 can always be evaluated on a smaller set
of CSFs before a full calculation is performed.
Therefore we always had control on the effects of the above mentioned
restrictions on the calculated values of $A$ and $Q$.

\section{Results}
\label{results}

\begin{figure}
  \includegraphics[width=.48\textwidth]{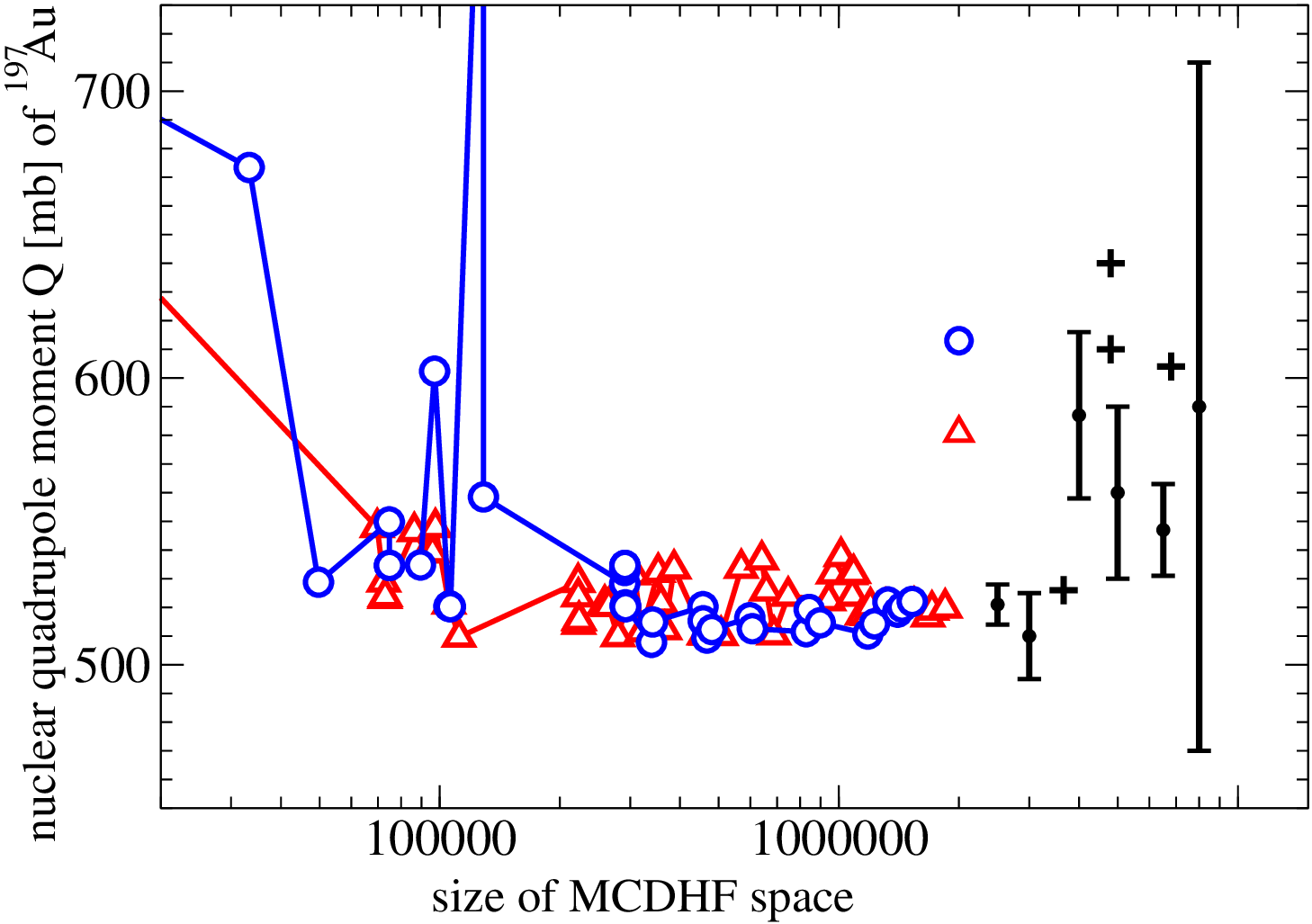}
\caption{(Color online)
         Nuclear quadrupole moment $Q$ [mb] of the $^{197}\!$Au isotope
 obtained from the calculated electric field gradients,
 as a function of
         the size of multiconfiguration expansions for the
         states 
$ 5d^9    6s^2 $~$ ^2 D _{3/2} $
 (triangles -- red online) and
$ 5d^9    6s^2 $~$ ^2 D _{5/2} $
 (circles -- blue online),
compared with other theoretical and experimental results.
%
The small corrections described in subsection~\ref{corrections}
are not included in the figure.
The values which represent multiconfiguration expansions
of sizes smaller
than 20000 are outside the figure,
except the uncorrelated Dirac-Hartree-Fock values, represented 
by the single triangle (red online)
for $ ^2 D _{3/2} $
and the single circle (blue online)
for $ ^2 D _{5/2} $.
The six values with error bars are from 
references~\cite{Yakobi:2007,Belpassi:2007,Itano:2006,PaladeWagner:2003,%
Powers:1974,ChildsGoodman:1966};
%
the four values without error bars, represented by pluses, are from 
references~\cite{ThierfelderSchwerdtfegerSaue:2007,Schwerdtfeger:2005,%
Blachman:1967};
all data are arranged in reverse chronological order, with the most recent
results
to the left.
}
\label{figureD3Q-D5Q-logarithmic}
\end{figure}


%
More extensive calculations
 turned out to be beyond the 100 node limit for this project on        
the Linux cluster at 
the National Institute of Standards and Technology (NIST), USA.
Therefore the calculations of the magnetic dipole constants
$A$ did not yield converged results.
%
%
As might be expected, the
effects of double and triple substitutions
are relatively larger for $A$ than for $Q$,
therefore the calculations of the $Q$ values were essentially converged;
they yield:
%
%
 Q($ ^2 D _{3/2} $)~=~519.829~mb, and
 Q($ ^2 D _{5/2} $)~=~522.066~mb, respectively.
%
%

\subsection{Corrections}
\label{corrections}


As mentioned in section~\ref{sdtq}, the
contributions arising from unrestricted
double substitutions from $4spdf$ orbitals
were evaluated separately and included additively in the final
$Q$ values.
They yield +0.312~mb and  +1.498~mb
for the two states
 $ ^2 D _{3/2} $ and
 $ ^2 D _{5/2} $, respectively.
%
%
The effects of the quadruple substitutions were also evaluated
separately, in a very limited fashion, and only for the 
 $ ^2 D _{3/2} $ state.
As explained in section~\ref{sdtq}, the correction arising from the 
quadruple substitutions for the $ ^2 D _{3/2} $ state lowers
the $Q$ value by 1.779~mb.
The dependence of the $Q$ values on double and triple
substitutions indicates that the quadruple correction might be
smaller for the $Q( ^2 D _{5/2})$ value
   than for the $Q( ^2 D _{3/2})$ value,
but we were unable to evaluate the former. Therefore
we assumed identical, $-$1.8~mb corrections for both states.
%
%
The corrections arising from the Breit interaction were 
calculated at the Dirac-Hartree-Fock level with full relaxation
i.e.~with a frequency-dependent Breit term
\begin{eqnarray}
\label{breit}
         B_{ij} & = & - \frac{\bm{\alpha}_{i} \cdot
                             \bm{\alpha}_{j}} {r_{ij}}
                      -\frac{\bm{\alpha}_{i} \cdot \bm{\alpha}_{j}}
         {r_{ij}} [\cos(\omega_{ij}r_{ij})-1]
         \nonumber \\
                &   & + c^2 (\bm{\alpha}_{i} \cdot
         \vec{\nabla}_{i})  (\bm{\alpha}_{j} \cdot
         \vec{\nabla}_{j})
         \frac{\cos(\omega_{ij}     r_{ij}/c) - 1}
                   {\omega_{ij}^{2} r_{ij}}
\end{eqnarray}
 included in the
self-consistent-field functional, using the \emph{mcdfgme} code \cite{desclaux:1993,indelicato:1995,mcdfgme}.
In the formula above,
$r_{ij}=\left|\vec{r}_{i}-\vec{r}_{j}\right|$ is the
 inter-electronic distance,
$\omega_{ij}$ is the energy of the photon
 exchanged between two electrons,
$\bm{\alpha}_{i}$ are Dirac matrices,
and $c=1/\alpha$ is the speed of light. 
The Breit corrections are highly  state-dependent
(see also~\cite{Yakobi:2007}, where the Gaunt part was evaluated)
%
       and yield 2.3~mb and 0.6~mb for the two states,
%
 $ ^2 D _{3/2} $ and
 $ ^2 D _{5/2} $, respectively.
The quantum electrodynamics (QED) corrections to the $Q$ values are expected to be very small.
We evaluated the VP (vacuum polarisation) correction with the \emph{mcdfgme} code, following Ref.~\cite{boucard:2000},
and obtained a value of the order of 0.01~\%.
%
%
%
When all above mentioned corrections are included,
the $Q$ values become:
Q($ ^2 D _{3/2} $)~=~520.641~mb and
Q($ ^2 D _{5/2} $)~=~522.364~mb.
The average of the above two results yields 
Q($ ^{197}\!$Au)~=~521.5~mb.

\subsection{Error estimate}
\label{error}

%
A rigorous, systematic treatment of the error bar of the calculated
electric quadrupole moment $Q$ would require
evaluation of the effects of: all omitted virtual orbitals,
all CSFs which were not included in the configuration expansions,
as well as all physical effects that were not included or
were treated approximately.
However, we were only able to obtain  very crude
estimates of certain sources of systematic errors.
We believe that 
none
exceeded 1~\%, but
the calculations presented in this
paper were far too extensive to permit a rigorous treatment
of the error. Therefore we have to resort to a
less rigorous method.

%
One of the frequently used methods of evaluation of the accuracy of calculated
electric quadrupole moments $Q$ is based on the simultaneous calculations
of magnetic dipole hyperfine constants $A$, and on subsequent
comparison of calculated $A$ values with their experimental
counterparts.
As mentioned above, the calculations of the
magnetic dipole constants
$A$ have not converged.
However, the amplitudes of the final oscillations of the two curves
representing the values of $A$ for the two states of interest
are comparable to the 
uncertainty of $A$ arising
from the accuracy  of the nuclear magnetic dipole moment value $\mu$.

There are currently two different  $\mu$ values in the
literature~\cite{Raghavan:1989,Stone:2005},
$\mu$ = 0.145746(9)  and  $\mu$ = 0.148158(8),
which
differ by about 2~\%.
Taken at face value, our results seem to favor
the smaller value, $\mu$~=~0.145746(9),
which, as mentioned in Sec.~\ref{theory},
has been used in the present calculations.
However, the overall accuracy of
our calculations
(in particular, the evaluation of higher-order terms)
does not permit us to draw a definitive conclusion.
Therefore, the difference between the two values of $\mu$
should rather be treated as a source of systematic error in the
determination of $A$.
Therefore, we did not push the calculations of magnetic dipole constants
$A$ further beyond their current level of convergence and, consequently,
the calculations of $A$ values could not be used as reliable sources
of error estimate for nuclear moments.

%
Another method
to estimate the accuracy of $Q$ is to consider the
differences between the final values obtained from different states.
However, in the present paper we were able to converge the calculations 
for only two atomic levels.
The difference between the results obtained for these two levels
turned out to be
quite small, which rendered this method useless in this
particular case.

%
Considering the computational methodology employed in this paper,
it is obvious that the final value depends on the choice of the
multiconfiguration expansions representing the last few points
on the curves in Fig.~\ref{figureD3Q-D5Q-logarithmic},
while the accuracy of the final value is connected with 
convergence of these curves.
Therefore, we based the estimate of the error bar
on the oscillations of the tail of
the two curves in
Fig.~\ref{figureD3Q-D5Q-logarithmic}.
The largest difference taken from 
the last few points on the curves representing
$ ^2 D _{3/2} $ and $ ^2 D _{5/2} $ states
amounts to 3~mb and 4~mb, respectively.
%
%
%
As an additional source of uncertainty we assumed the additive
corrections described in subsection~\ref{corrections},
since all of them were evaluated in a rather crude approximation.
%
For instance,
the contribution of the Breit 
interaction was calculated at the Dirac-Hartree-Fock level, without
regard for electron correlation effects. 
%
%
When all above sources of uncertainly are taken into account
the total error bar amounts to 5~mb, which yields our
final calculated value of quadrupole moment
Q($^{197}\!$Au)~=~521.5~$\pm$~{5.0}~mb.

\section{Comparisons}
\label{comparisons}

The results of our calculation are compared with previous evaluations
in Table~\ref{Qsummary} and
 in Fig.~\ref{figureD3Q-D5Q-logarithmic}.
It is worth noting that our result is in agreement with three most
recent theoretical values,
obtained with three different methods,
but all these recent results (including ours)
are considerably smaller than other, earlier values.

%
%
Yakobi~{\sl et al}~\cite{Yakobi:2007}
performed calculations for the 
$ 5d^9    6s^2 $~$ ^2 D _{3/2} $ and
$ 5d^9    6s^2 $~$ ^2 D _{5/2} $ states of atomic gold
within the four-component Dirac-Coulomb
framework~\cite{Eliav:1994,Kaldor:1998}.
They correlated 51 out of the 79 electrons 
in the large basis sets (up to 26s22p18d12f8g5h uncontracted
Gaussian functions)
with the relativistic Fock-space coupled cluster
method, including single and double excitations (CCSD).
The contribution of the Gaunt term, the main part of
the Breit interaction, was also evaluated.

%
%
Belpassi~{\sl et al}~\cite{Belpassi:2007}
performed
molecular relativistic Dirac-Coulomb-Gaunt Hartree-Fock
calculations~\cite{Pernpointner:2001}
for a series of molecules: AuF, XeAuF, KrAuF, ArAuF, (OC)AuF, and AuH.
The electronic correlation contributions were included at
CCSD(T) and CCSD-T levels.
The value of the nuclear quadrupole moment $Q$ was obtained from
the determinations of the electric field gradient at the gold nucleus
for the above mentioned molecules,
combined with experimental values of the nuclear quadrupole coupling constants.

%
%
Thierfelder~{\sl et al}~\cite{ThierfelderSchwerdtfegerSaue:2007}
performed four-component relativistic density-functional (DFT)
calculations 
for diatomic compounds CuX and AuX
(X = H, F, Cl, Br, and I)
with and without CO attached, i.e.,
OC-CuX and OC-AuX (X = F, Cl, Br, and I).
They employed a newly developed
functional~\cite{Yanai:2004}, whose role is to correctly
describe the long-range part of exchange interactions~\cite{Goll:2007},
and obtained the averaged result~Q~=~526~mb. This value is within the error
bounds of our value.

%
Our result, in turn, falls
within the error bounds published by
Belpassi~{\sl et al} (Q~=~510(15)~mb),
%
as well as those by
Yakobi~{\sl et al} (Q~=~521(7)~mb).
%
%
The agreement with 
Yakobi~{\sl et al} may be somewhat
accidental because particular contributions show larger differences.
The two outstanding differences arise from triple substitutions
and from deep core orbitals. 
Yakobi~{\sl et al} evaluated the effect of the triple substitutions
by performing single-reference CCSD(T) calculation
for the  $ ^2 D _{5/2} $ level,
and obtained a 0.3~\% shift.
The effect of triple substitutions is indeed smaller
for the $ ^2 D _{5/2} $ level,
but
for the $ ^2 D _{3/2} $ level our calculations indicate
   a shift of the order of 2~\%.
However, this discrepancy may be attributed to the methodological
 differences in the two papers.
The definition of triple substitutions in the
configuration-interaction method used here
differs
substantially from that in the CCSD(T) approach,
due to the exponential nature of the coupled-cluster operator.
The coupled-cluster  approximation includes a subset of the CI triple
substitutions (the "unlinked" diagrams),
as well as that of higher order substitutions,
already at the CCSD level.
The CCSD(T) yields only the "linked" part as the effect of
the triple substitutions.
Therefore,
the contribution of the
 CI triple substitutions may indeed be expected to be         
larger than that of the CC triple substitutions.

Another difference arises from contributions of
deep core orbitals. 
The
effects of $3spd$, $2sp$, and $1s$ orbitals
were neglected by Yakobi~{\sl et al}, while in our calculations
they were all included. Their combined effect was to lower the
 $Q$ value by about 2~\%.

%
\begin{table}
\caption{Comparison of the present $Q$($^{197}$Au) value
(in~mb~=~10$^{-31}$~m$^2$) with other recent values
and with previous (muonic) standard value.
}
\label{Qsummary}
\begin{tabular}{llll}
\colrule
Reference  &  Source  &  Q($^{197} \! $Au)  &  \\ 
\colrule
This work & Au atom, $^2D_{3/2}$, $^2D_{5/2}$ &
  \multicolumn{1}{l}{ 521.5$\pm$5.0 } \\
%
Yakobi et al~\cite{Yakobi:2007} & Au atom, $^2D_{3/2}$, $^2D_{5/2}$ &
  \multicolumn{1}{l}{ 521$\pm$7 } \\
%
Belpassi et al~\cite{Belpassi:2007} & AuF, LAuF molecules  &
  \multicolumn{1}{l}{ 510$\pm$15 } \\
Thierfelder et al~\cite{ThierfelderSchwerdtfegerSaue:2007} \,\,\, &
  AuX, LAuX molecules \,\,\,\,\, &
  \multicolumn{1}{l}{ 526 } \\
%
Powers et al~\cite{Powers:1974} & Muonic &
  \multicolumn{1}{l}{ 547$\pm$16 } \\
%
\colrule
\end{tabular}
\end{table}
%

\section*{Conclusions}
\label{conclusions}

The multiconfiguration Dirac-Hartree-Fock (MCDHF) model has been employed
to calculate the expectation values
responsible for the hyperfine splittings of the
$ 5d^9    6s^2 $~$ ^2 D _{3/2} $ and
$ 5d^9    6s^2 $~$ ^2 D _{5/2} $
levels of atomic gold.
To our knowledge, this  is the first calculation in which
\mbox{one-,} two-, and three-body electron correlation effects were included
and saturated for electric quadrupole hyperfine values 
of a heavy, open-shell, neutral atom.
The correlation effects involving all 79 electrons were accounted for
with a procedure  that  is             equivalent to a full 
 Complete Active Space calculation.
All electron correlation effects were explicitly accounted
for at a 1~\% level of precision or better.
Calculated electric field gradients, together with experimental
values of the electric quadrupole hyperfine structure constants $B$,
allow us to extract a nuclear electric quadrupole moment
 $Q$~=~521.5(5.0)~mb
of $^{197}$Au.
If taken at face value, the summary in table~\ref{Qsummary}
suggests that our
$Q$ value, together with that of Yakobi~{\sl et al}~\cite{Yakobi:2007},
could be the new standard value.

\vspace{1cm}

\section*{Acknowledgements}
%
%
This work was supported by
the Polish Ministry of Science and Higher Education (MNiSW)
in the framework of scientific grant No.~1~P03B~110~30
awarded for the years 2006-2009.
PJ acknowledges support from the
Swedish Research Council (Vetenskapsr{\aa}det).
PP belongs to the Finnish Center of Excellence
in Computational Molecular Science (CMS).
The visits of JB at Helsinki were supported by The Academy of Finland.
The large scale calculations were performed on 
the Raritan Linux cluster at 
the National Institute of Standards and Technology (NIST)
in Gaithersburg, USA.
JB would like to express his gratitude for the hospitality
which was extended to him during his visits
to the Chemistry Department of the University of Helsinki
and the Atomic Spectroscopy Group at NIST.
PJ acknowledges the support of the                                   
Helmholtz Alliance Program of the Helmholtz Association, contract               
HA-216 ``Extremes of Density and Temperature: Cosmic Matter in the              
 Laboratory''.
Laboratoire Kastler Brossel is ``Unit\'e Mixte de Recherche du CNRS, de
l' ENS et de l'UPMC n$^{\circ}$ 8552''.
%
We thank the (anonymous) referee for pointing our attention to the
structural differences between CI and CC methods.


\bibliography{xet}

\end{document}